\documentclass[aps,prl,twocolumn,showpacs,unsortedaddress,superscriptaddress]{revtex4-1}

\usepackage{hyperref}
\usepackage{epsfig}
\usepackage{graphicx}
\usepackage{subfigure}
\usepackage{latexsym}
\usepackage{color}
\usepackage{fullpage}
\usepackage{dcolumn}
\usepackage{bm}
\usepackage{ulem}
\usepackage{units}
\usepackage{amsmath}

\usepackage{units}

\begin{document}

\title{Degenerate epitaxy-driven defects in monolayer silicon oxide onto ruthenium}

\author{Shashank Mathur}
\affiliation{Univ. Grenoble Alpes, F-38000 Grenoble, France}
\affiliation{CNRS, Inst NEEL, F-38042 Grenoble, France}
\affiliation{CEA, INAC-SP2M, Grenoble, F-38054, France}

\author{Sergio Vlaic}
\affiliation{Univ. Grenoble Alpes, F-38000 Grenoble, France}
\affiliation{CNRS, Inst NEEL, F-38042 Grenoble, France}
\affiliation{LPEM-UMR8213/CNRS-ESPCI ParisTech-UPMC, 10 rue Vauquelin, 75005 Paris}

\author{Eduardo Machado-Charry}
\affiliation{Univ. Grenoble Alpes, F-38000 Grenoble, France}
\affiliation{CEA, INAC-SP2M, Grenoble, F-38054, France}

\author{Anh-Duc Vu}
\affiliation{Univ. Grenoble Alpes, F-38000 Grenoble, France}
\affiliation{CNRS, Inst NEEL, F-38042 Grenoble, France}

\author{Val\'{e}rie Guisset}
\affiliation{Univ. Grenoble Alpes, F-38000 Grenoble, France}
\affiliation{CNRS, Inst NEEL, F-38042 Grenoble, France}

\author{Philippe David}
\affiliation{Univ. Grenoble Alpes, F-38000 Grenoble, France}
\affiliation{CNRS, Inst NEEL, F-38042 Grenoble, France}

\author{Emmanuel Hadji}
\affiliation{Univ. Grenoble Alpes, F-38000 Grenoble, France}
\affiliation{CEA, INAC-SP2M, Grenoble, F-38054, France}

\author{Pascal Pochet}
\affiliation{Univ. Grenoble Alpes, F-38000 Grenoble, France}
\affiliation{CEA, INAC-SP2M, Grenoble, F-38054, France}
\email{pascal.pochet@cea.fr}

\author{Johann Coraux}
\affiliation{Univ. Grenoble Alpes, F-38000 Grenoble, France}
\affiliation{CNRS, Inst NEEL, F-38042 Grenoble, France}
\email{johann.coraux@neel.cnrs.fr}

\date{\today}%

\begin{abstract}

The structure of the ultimately-thin crystalline allotrope of silicon oxide, prepared onto a ruthenium surface, is unveiled down to atomic scale with chemical sensitivity, thanks to high resolution scanning tunneling microscopy and first principle calculations. An ordered oxygen lattice is imaged which coexists with the two-dimensional monolayer oxide. This coexistence signals a displacive transformation from an oxygen reconstructed-Ru(0001) to silicon oxide, along which latterally-shifted domains form, each with equivalent and degenerate epitaxial relationships with the substrate. The unavoidable character of defects at boundaries between these domains appeals for the development of alternative methods capable of producing single-crystalline two-dimensional oxides.\\
\\
See \href{http://dx.doi.org/10.1103/PhysRevB.92.161410}{\textcolor{blue}{\underline{Phys. Rev. B 92, 161410(R) (2015)}}} for published version of this work.

\end{abstract}

\maketitle

Silicon oxide is a widely abundant compound existing in various forms, from amorphous to crystalline, from bulk to porous and thin films. It is a dielectric material widely used as a thin film for capacitive control of charge carriers in microelectronics' conductive channels \cite{Atalla1962}. It is also a widespread porous support for catalysis, efficiently dispersing nanoparticles and playing active role \cite{Morrow1976,Inaki2002,Groppo2005} in the catalytic reactions. Its ill-defined amorphous and complex three-dimensional structure however hinders the understanding of the elementary processes driving the overall performance. Deeper insights are expected if resorting to ultrathin silicon oxide films of well-defined structure, whose defect nature and density can be controlled, which are amenable to high resolution surface-sensitive probes, as was shown for various other oxides \cite{Freund2008}. Crystalline oxide thin films are also appealing as building block in van der Waals heterostructures with ultimately-thin two-dimensional (2D) crystals (\textit{e.g.} graphene, silicene, transition metal dichalcogenides) \cite{Britnell2013}. Such devices, already demonstrated at small scales with boron nitride flakes as decoupling layers \cite{Dean2010,Britnell2012}, are within reach with larger area ultrathin silicon oxide films, which can be grown onto graphene \cite{Huang2012} and serve as an efficient decoupling layer between graphene and a metal \cite{Lizzit2012}.

The thinnest crystalline allotrope of silicon oxide is a monolayer (ML) of corner-sharing SiO$_4$ tetrahedra forming a honeycomb lattice on metal surfaces like Mo(112) \cite{He1992,Luo2001,Schroeder2000} and Ru(0001) \cite{Yang2012}. The (0001) surface of ruthenium, whose crystal symmetry is similar to that of the oxide, and whose lattice parameter is close to half the oxide's, is expected to stabilize the latter phase \cite{BenRomdhane2013} without stress-relief defects such as misfit dislocations. Besides, the strong ruthenium-oxygen interaction prevents in-plane disorientations between ML silicon oxide and Ru(0001), leading to a single crystallographic orientation \cite{Yang2012}. Determining the precise nature of the bonding and structure imposed by this interaction is required to rationalize growth, defect formation, and chemical reactivity. Such information is the bottleneck to the control of the catalytic activity and large-area high-quality preparation of ultimately-thin crystalline oxides.

Local microscopies and ensemble-averaged spectroscopies have provided insights into the structure of ML silicon oxide. The former revealed a honeycomb lattice \cite{Yang2012} with a thickness compatible with Ru-O-Si segments standing perpendicular to the surface \cite{BenRomdhane2013}. The latter detected oxygen atoms involved in different kinds of bonds with silicon and ruthenium \cite{Yang2012}. A model accounting for these observations has been proposed which is stable according to first principle calculations \cite{Yang2012,BenRomdhane2013}. In the absence of truly atomic scale imaging with chemical sensitivity, this model could not yet be further confirmed however. In this work we solve this issue by employing scanning tunneling microscopy (STM) and density functional theory (DFT) calculations. We find compelling evidence by direct imaging that ML silicon oxide binds to the metal through oxygen atoms, half of them directly atop ruthenium atoms, the other half in $fcc$ (face-centered cubic) hollow sites of ruthenium. In addition, we provide clearcut direct imaging of a (2$\times$2) oxygen reconstruction of Ru(0001) coexisting with ML silicon oxide. This coexistence, which has been conjectured in several reports, yet whose origin is debated, is here given grounds in light of a displacive transformation. The transformation, of an ordered oxygen reconstruction exposed to silicon into the oxide, is found to be the main source of defects in the latter, as revealed by STM and reflection high-energy electron diffraction (RHEED). The defects, chains of heptagons and pentagons unexpectedly having preferential orientation, ensure continuity of the atomic lattice across laterally-shifted domains. The coexistence of these domains, in other words degenerate epitaxy, translates the larger number (three \textit{vs} one) of lattices which ML silicon oxide comprises, compared to Ru(0001). It represents a serious hurdle to the preparation of defect-free ML silicon oxide.

\begin{figure*}[hbt]
  \begin{center}
  \includegraphics[width=160mm]{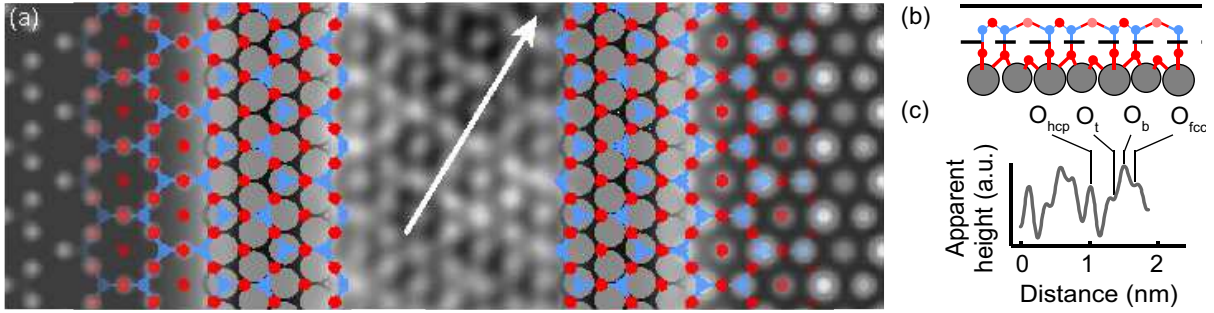}
  \caption{\label{fig1}(Color online) (a) STM topograph (1.8$\times$2.4~nm$^2$, center, $I_\mathrm{t}$ = 2~nA, $U_\mathrm{t}$ = 0.9~V), overlaid with a ball-model of ML silicon oxide and two constant heigh PEDM cuts, deduced from DFT calculations, closer (right, 2.2~\AA) and further (left, 3.5~\AA) from the surface. (b) Side view DFT-calculated model of ML silicon oxide along the white arrow shown in (a), and (c) corresponding apparent height profile. The full and dashed vertical lines in (b) mark the heights at which the top and bottom (respectively) constant height cuts in (a) have been calculated.}
  \end{center}
\end{figure*}

Monolayer silicon oxide was prepared under ultra-high vacuum by first exposing the clean Ru(0001) to an oxygen dose ensuring the formation of a fully developed (2$\times$2)-3O reconstruction of oxygen atoms (see Supplemental Material \cite{noteSI}). This phase protects Ru(0001) from silicide formation in the next step, which consists in evaporating silicon onto the surface and subsequently annealing the sample \cite{Shaikhutdinov2013} (see Supplemental Material \cite{noteSI}).

Direct imaging of ML silicon oxide on Ru(0001) with high resolution is scarse. To our knowledge only STM has been employed for this purpose, and was performed in constant tunneling current ($I_\mathrm{t}$) mode with a high tunneling resistance (ratio between the tunneling bias, $U_\mathrm{t}$, and $I_\mathrm{t}$), of the order of 10~M$\Omega$ \cite{Yang2013,Lichtenstein2012}. In such conditions the hexagonal lattice of ML silicon oxide (of 5.4~\AA-period) was unveiled, but finer details about the arrangement of atoms could not be obtained. In our STM experiments, we used a tunnelling resistance one order of magnitude smaller, which imposes smaller tip-sample distances. In such conditions, and after extensive tip preparation presumably yielding, occasionally, very sharp tip apex, we achieve high spatial resolution with chemical information (Figs.~\ref{fig1}a,c). Each hexagon of the honeycomb lattice displays a height modulation, with the corners lower (darker) than the (bright) middle of each of the segments. Interestingly, half of the corners appear lower than the other half, according to a three-fold symmetry. Besides, each hexagon presents a (bright) protrusion at its center.

In order to relate these dark and bright features to the atomic arrangement in ML silicon oxide, we performed DFT calculations (see Supplemental Material \cite{noteSI}) of the partial electron density maps (PEDMs) for the predicted most stable configuration. In this configuration, shown in Figs.~\ref{fig1}a,b, the oxide bonds Ru(0001) \textit{via} six oxygen atoms, half of them sitting directly on top of a Ru atom (O$_\mathrm{t}$), half of them sitting onto a $fcc$ hollow site of Ru(0001) (O$_\mathrm{fcc}$) \cite{Yang2012,BenRomdhane2013}. Besides, an isolated oxygen atom occupies a $hcp$ (hexagonal close-packed) hollow site of Ru(0001) (O$_\mathrm{fcc}$), whose position is the center of the hexagon defined by the Si-O-Si segments (the oxygen atoms in these segments, which bridge two silicon atoms, are referred to as O$_\mathrm{b}$). 

\begin{figure}[hbt]
  \begin{center}
  \includegraphics[width=71.1mm]{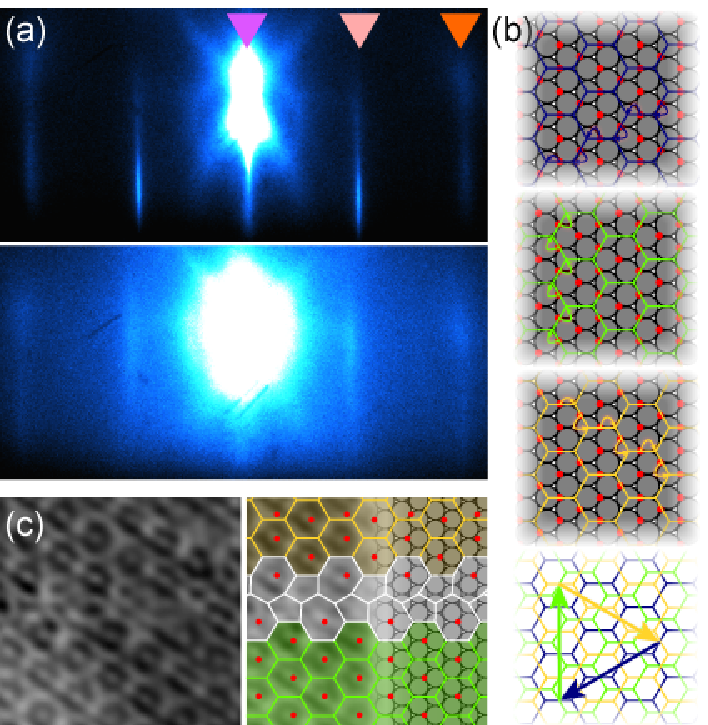}
  \caption{\label{fig2}(Color online) (a) RHEED pattern for the (2$\times$2)-3O reconstruction prior (top) and after (bottom) deposition of 1.4$\pm$0.2~ML Si. The three downward-pointing triangles in each RHEED pattern mark the positions of the specularly-reflected beam (violet triangle) and of the reconstruction (pale orange triangle) and Ru (darker orange triangle) diffraction streaks. (b) Ball-models showing the three equivalent (displacive) transformations of oxygen atoms in the (2$\times$2)-3O reconstruction, into the lattice of O atoms in ML silicon oxide which are in direct contact with Ru(0001). The bottom pannel highlights the relative lateral shifts of the three corresponding lattices of ML silicon oxide. (c) STM topograph (left, 3$\times$3~nm$^2$, $I_\mathrm{t}$ = 2~nA, $U_\mathrm{t}$ = 0.9~V) revealing a boundary parallel to an armchair direction of ML silicon oxide, between antiphase domains which are shifted along two different $\left\langle1\bar{1}00\right\rangle$ directions (corresponding to the yellow and blue directions shown in (b)) with respect to the (2$\times$2)-3O reconstruction, and same image overlaid with a ball model (right pannel).}
  \end{center}
\end{figure}

We computed a constant height PEDM cut 3.5~\AA\, above the Ru(0001) surface (left pannel of Fig.~\ref{fig1}a), \textit{i.e.} above the ML silicon oxide surface. A prominent contribution is observed at the location of the O$_\mathrm{b}$ sites, where actually oxygen atoms terminate the surface (Fig.~\ref{fig1}b). On the contrary, no or negligible contribution is observed at the O$_\mathrm{fcc}$, O$_\mathrm{hcp}$, and O$_\mathrm{t}$ sites. At the corresponding positions in the STM image, the apparent height is about two thirds of that of the O$_\mathrm{b}$ sites. At O$_\mathrm{fcc}$ and O$_\mathrm{t}$ sites, we ascribe the lower apparent height to a weaker local density of states around the silicon atoms, which sit atop these sites, presumably in reason of strong charge transfers towards the underlying oxygen atoms. At the O$_\mathrm{hcp}$ sites, an oxygen atom is actually directly accessible to the STM tip. Assuming for simplicity that the local electronic density is comparable around these oxygen atoms and around those on O$_\mathrm{b}$ sites, the observed lower apparent height is ascribed to a true heigh difference -- the STM feedback loop approaches the tip to the surface in order to maintain a constant tunneling current. A PEDM cut closer to the surface, at 2.2~\AA\, (right pannel of Fig.~\ref{fig1}a), was hence calculated in order to account for the lower STM tip position in the center of the hexagonal rings of ML silicon oxide. As expected bright protrusions, comparable to those observed at 3.5~\AA\, height, are found where the oxygen atoms bond to Ru(0001), at O$_\mathrm{fcc}$ sites. The overall qualitative agreement between the PEDM cuts and STM images confirms for the first time by real space imaging the relevance of the model proposed \cite{Yang2012} on the basis of energetic considerations (DFT results) and chemical and vibrational signatures of oxygen atoms involved in various kinds of bonds with Ru and/or Si (vibrational and photoelectron spectroscopy). Additional experimental data further supporting the structural model is obtained at the vicinity of vacancies in ML silicon oxide, is shown in Supplemental Material \cite{noteSI}.

The coexistence of ML silicon oxide with a (2$\times$2) oxygen reconstruction formed by the O$_\mathrm{hcp}$ (ring centers) atoms was argued to stabilize the system, yet marginally, below 0.1\% of its total energy \cite{BenRomdhane2013}. This prediction suggests that silicon oxide formation does not strongly promote further chemisorption of oxygen, but that rather, the (2$\times$2) oxygen reconstruction is inherited from the passivating (2$\times$2)-3O reconstruction. Such a relationship between ML silicon oxide and the oxygen reconstruction is in line with the domains of the former coexisting with the latter for partial coverage \cite{Lichtenstein2012}. A preservation, at least partial, of atomic order after silicon deposition onto the (2$\times$2)-3O-reconstructed Ru(0001) surface (the subsequent step yielding fully-developed ML silicon oxide consisting in thermal annealing) should occur in this case, which should show up in diffraction. RHEED patterns indeed show a set of diffraction streaks halfway between the Ru ones and the specularly reflected beam, in addition to a strong diffuse scattering background. These streaks signal an ordered structure with a lattice parameter twice that of Ru(0001), and are also found for the pure (2$\times$2)-3O reconstruction, while diffuse scattering points to disorder (compare the two panels of Fig.~\ref{fig2}a). \textit{In operando} monitoring with RHEED all along the growth process (data not shown here) reveals the persistance of the streaks even in presence of strong diffuse scattering background, ever since the formation of the oxygen reconstruction. Detailed analysis of this data, complemented for instance with high resolution STM imaging after annealing at increasing temperature, both beyond the scope of the present work, would help to elucidate the exact nature of this mixed ordered-disordered phase. At this point we surmise that oxygen atoms preserve their preferencial bonding along a (2$\times$2)-3O reconstruction, and that silicon atoms bind to them in a disordered fashion.

A geometrical operation transforming the lattice of oxygen atoms in direct contact to ruthenium, from the (2$\times$2)-3O reconstruction to ML silicon oxide, is shown in Fig.~\ref{fig2}b. It involves the shortest possible displacements of only two of the three (2$\times$2) sub-lattices constituting the (2$\times$2)-3O reconstruction: the first sub-lattice moves from the $hcp$ to a nearest $fcc$ site, while the second one moves from the $hcp$ to a nearest atop site. Based on the above, we propose that this transformation is not simply a geometric one, but the essence of the transformation from the mixed ordered-disordered phase, discussed in the previous paragraph, to the ordered ML silicon oxide obtained after the 850$^\circ$C annealing. If such a transformation exists, it must be displacive, \textit{i.e.} it must involve a (small) global shift of part of the oxygen lattice, as we argue latter. In this case, the transformation is three-fold degenerate since it can occur in the three equivalent, 120$^\circ$-rotated, $\left\langle1\bar{1}00\right\rangle$ directions of the Ru(0001) surface (zigzag directions of silicon oxide), depending on which of the three oxygen sub-lattices remains unchanged.

\begin{figure}[hbt]
  \begin{center}
  \includegraphics[width=79.6mm]{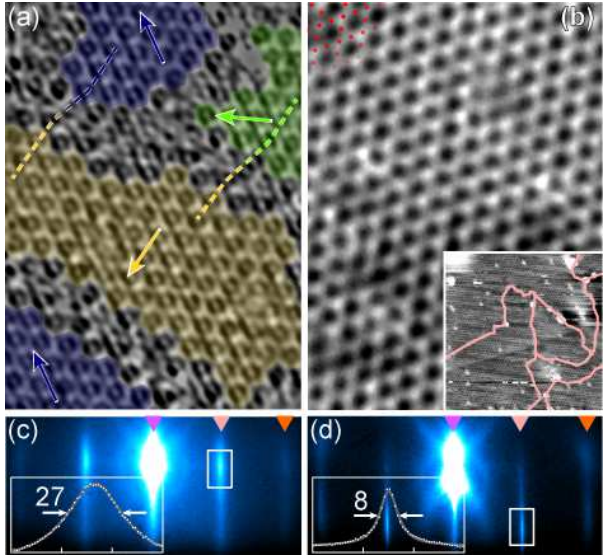}
  \caption{\label{fig3}(Color online) (a) STM topograph (8.5$\times$6.2~nm$^2$, $I_\mathrm{t}$ = 2~nA, $U_\mathrm{t}$ = 0.9~V) of ML silicon oxide comprising antiphase domains. The domains are colored according to the code used for Fig.~\ref{fig2}b; The domains are laterally shifted, \textit{i.e.} a straight line crossing the boundary cannot be drawn that goes through the same crystallographic sites in two adjacent domains (see dashed lines). (b) STM topograph ($I_\mathrm{t}$ = 1.9~nA, $U_\mathrm{t}$ = 2~V), of the same size as (a), of the (2$\times$2)-3O reconstruction. Inset: larger scale STM topograph (25$\times$25~nm$^2$, $I_\mathrm{t}$ = 1.8~nA, $U_\mathrm{t}$ = 2.1~V) overlaid with contours of antiphase domains. (c,d) RHEED patterns for ML silicon oxide (c) and the (2$\times$2)-3O reconstruction (d). Insets: intensity profiles averaged in the horizontal direction in the white frames; data is shown as white contour-dots, gray solid lines are gaussian fits to the data; full-widths at half maximum are indicated in pixel units (27 and 8).}
  \end{center}
\end{figure}

Although the crystallographic orientation of ML silicon oxide is unique and no rotational domains are found, the degeneracy of the displacive transformation imposes the formation of domains whose registries with the Ru(0001) substrate are laterally shifted one with respect to the other, by ca. 2.7~\AA\, (the nearest neighbor distance in Ru(0001)) along $[1\bar{1}00]$ (Fig.~\ref{fig2}c). These so-called antiphase domains are a typical instance of degenerate epitaxy \cite{Chan1994}, frequently encountered for ultrathin oxides prepared on metal surfaces \cite{Kaya2007,Tao2011,Konig2011,Gragnaniello2011,Boscoboinik2013} as a consequence of their structure consisting of several laterally shifted sub-lattices. Between the three kinds of domains, we almost exclusively observe (Fig.~\ref{fig3}a) antiphase domain boundaries aligned with armchair directions of silicon oxide ($\left\langle11\bar{2}0\right\rangle$ of Ru(0001)), consistent with previously published STM data \cite{Yang2013}. As expected for a boundary accommodating a perpendicular misalignement of atomic rows in a honeycomb lattice, polygonal rings with an odd number of segments, at least one which of which can orient along directions between $\left\langle11\bar{2}0\right\rangle$ and $\left\langle1\bar{1}00\right\rangle$, are found. The nature of the polygons constituting the boundaries, pairs of heptagons and pentagons \cite{Yang2013}, is determined by carefully inspecting the complex STM topography associated to distinctive apparent heights, which are for instance expected for rings' corners sitting on $fcc$ and atop sites (Fig.~\ref{fig2}c). An oxygen atom is observed onto a $hcp$ hollow site of Ru(0001) inside the heptagons (Fig.~\ref{fig2}c). Inside pentagons we do not observe an oxygen atom, which could signal steric hindrance of either the STM tip or of the oxygen atoms themselves due to the small size of the ring.

Noteworthy, we do not observe other kinds of domain boundaries (see Supplemental Material \cite{noteSI}), for instance those which align zigzag directions. Why these kinds of boundaries do not form could be a sign of their higher energetic cost and/or the cost of the corresponding bond breaking/re-arrangement for their formation.

Antiphase boundaries (and degenerate epitaxy) not only are found in ML silicon oxide, but also in the (2$\times$2)-3O reconstruction. In the latter case, their surface density can be tuned across several orders of magnitude depending on the oxygen deposition temperature and subsequent annealing (see Fig.~\ref{fig3}b and Supplemental Material \cite{noteSI}), which control oxygen atom's surface diffusion \cite{Wintterlin1997,Hammer2000}. We find that the size of the (2$\times$2)-3O antiphase domains matches that of the domains of the parent (2$\times$1) oxygen reconstruction from which it evolves. This observation points to a surface mobility of oxygen atoms at least as high in the (2$\times$1) as in the (2$\times$2), indicating that close-by chemisorbed oxygen atom do not hinder oxygen surface diffusion as observed in case of oxygen dimers onto Ru(0001) \cite{Renisch1999}. While (2$\times$2)-3O domains of size of several 10~nm are achieved by oxygen deposition at 350$^\circ$C followed by annealing at 850$^\circ$C (inset of Fig.~\ref{fig3}b), the domain size in ML silicon oxide, obtained after room temperature deposition of silicon onto this reconstructed surface followed by 850$^\circ$C annealing, is an order of magnitude lower (see Supplemental Material \cite{noteSI}). The lower crystallinity of ML silicon oxide is evident when comparing its diffraction pattern with that of the oxygen reconstruction. Even if care should be taken in extracting quantitative information from RHEED patterns due to complex electron diffraction effects, the corresponding diffraction streaks indeed show a substantially stronger finite-size broadening in the former case (Figs.~\ref{fig3}c,d).

The proposed transformation, from the (2$\times$2)-3O reconstruction to ML silicon oxide, breaks individual domains of the oxygen reconstruction into the three kinds of domains separated by the antiphase boundaries that we discussed in a previous paragraph -- which supports the above claim of a displacive transformation. This transformation is the main source of defects (antiphase boundaries) in ML silicon oxide. We exclude heteroepitaxial stress between Ru(0001) and the oxide as the driving force for antiphase domain formation. The minimal size of oxide domains in such a case, as assessed by assuming that the ca. 2\% lattice mismatch \cite{BenRomdhane2013} between Ru(0001) and ML silicon oxide is fully accommodated by insertion of antiphase boundaries, would indeed be of about 30 nm, 	\textit{i.e.} already several times larger the domain size observed by STM.

In summary, we establish, by imaging and first principles simulations, the so-far experimentally unresolved atomic structure of ML silicon oxide on Ru(0001) and the precise nature of the bonding between the two materials. The oxide is found to coexist with a (2$\times$2) reconstruction of oxygen atoms inside the rings of silicon oxide. We propose that the transition from (2$\times$2)-3O-reconstructed Ru(0001) to silicon oxide consists in a displacive transformation, which is degenerate and yields antiphase boundaries that are exclusively oriented along armchair directions and consist of pairs of heptagons and pentagons. Such a transformation is the main source of defects in ML silicon oxide and challenges the production of high quality samples. Understanding the influence of the defects on the chemical and physical properties calls for dedicated studies. Even if not yet discussed in the context of 2D crystals, we anticipate that antiphase boundaries should also be found in, \textit{e.g.}, single-layer graphene and transition metal dichalocogenides, especially in case of strong interaction between the 2D crystal and the substrate.

\begin{acknowledgments}
Financial support from Agence Nationale de la Recherche through the ANR-12-BS-1000-401-NANOCELLS and ANR-14-OHRI-0004-01 2DTransformers contracts is acknoledged. S.V. acknowledges financial support from the Swiss National Science Foundation through project No. PBELP2-146587. S.M. acknowledges support from the LANEF LABEX. P.P and E.H. acknowledge the financial support of the Nanoscience CEA program. The DFT calculations were done using French supercomputers (GENCI) through project 6194.
\end{acknowledgments}


%

\end{document}